\documentclass[twocolumn,pre,twoside,superscriptaddress,floatfix]{revtex4-1}

\usepackage{amsmath}
\usepackage{amssymb}
\usepackage{graphicx}
\usepackage{dcolumn}
\usepackage{bm}
\usepackage{xcolor}

\def\dps{\displaystyle}
\def\m#1{\mathrm{#1}}
\def\d{\mathrm{d}}
\def\p{\partial}

\def\ol#1{\overline{#1}}

\def\wh#1{\widehat{#1}}
\def\wt#1{\widetilde{#1}}
\def\vec#1{\mathbf{#1}}

\def\sinc{\mathop\mathrm{sinc}}

\begin{document}


\title{Structure of electrolyte solutions at non-uniformly charged
surfaces on a variety of length scales}

\author{Markus Bier}
\email{markus.bier@fhws.de}
\affiliation
{
   Max-Planck-Institut f\"{u}r Intelligente Systeme,\\ 
   Heisenbergstr.\ 3,
   70569 Stuttgart,
   Germany
}
\affiliation   
{
   Institut f\"{u}r Theoretische Physik IV,
   Universit\"{a}t Stuttgart,
   Pfaffenwaldring 57,
   70569 Stuttgart,
   Germany
}
\affiliation
{
   Fakult\"{a}t Angewandte Natur- und Geisteswissenschaften,
   Hochschule f\"{u}r angewandte Wissenschaften W\"{u}rzburg-Schweinfurt,
   Ignaz-Sch\"{o}n-Str.\ 11, 97421 Schweinfurt, Germany
}
\author{Maximilian Mu{\ss}otter}
\affiliation
{
   Max-Planck-Institut f\"{u}r Intelligente Systeme,\\ 
   Heisenbergstr.\ 3,
   70569 Stuttgart,
   Germany
}
\affiliation
{
   Institut f\"{u}r Theoretische Physik IV,
   Universit\"{a}t Stuttgart,
   Pfaffenwaldring 57,
   70569 Stuttgart,
   Germany
}
\author{S.\ Dietrich}
\affiliation
{
   Max-Planck-Institut f\"{u}r Intelligente Systeme,\\ 
   Heisenbergstr.\ 3,
   70569 Stuttgart,
   Germany
}
\affiliation
{
   Institut f\"{u}r Theoretische Physik IV,
   Universit\"{a}t Stuttgart,
   Pfaffenwaldring 57,
   70569 Stuttgart,
   Germany
}

\date{25 August 2022}

\begin{abstract}
The structures of dilute electrolyte solutions close to non-uniformly
charged planar substrates are systematically studied within the
entire spectrum of microscopic to macroscopic length scales by means
of a unified classical density functional theory (DFT) approach.
This is in contrast to previous investigations, which are applicable either
to short or to long length scales.
It turns out that interactions with microscopic ranges, e.g., due to
the hard cores of the fluid molecules and ions, have negligible influence
on the formation of non-uniform lateral structures of the electrolyte
solutions.
This partly justifies the Debye-H\"{u}ckel approximation schemes applied
in previous studies of that system.
In general, a coupling between the lateral and the normal fluid structures
leads to the phenomenology that, upon increasing the distance from the
substrate, less details of the lateral non-uniformities contribute to
the fluid structure, such that ultimately only large-scale surface features
remain relevant.
It can be expected that this picture also applies to other fluids
characterized by several length scales.
\end{abstract}

\maketitle


\section{\label{sec:intro}Introduction}

Historically, the theoretical study of solid-fluid interfaces has naturally
started with the investigation of idealized surfaces with laterally uniform
properties \cite{Gibbs,Helmholtz1879,Gouy1909,Gouy1910,Chapman1913,
Grahame1947} instead of realistic models of surfaces with geometrical,
chemical, or electrical non-uniformities.
This approach was justified, on the one hand, by the initial lack of knowledge
about the microscopic structure of real surfaces, and, on the other hand, by
the computational advantages gained from exploiting lateral symmetries.
However, in particular in the context of electrochemistry and colloidal
science, efforts have been made to include surface non-uniformities into
the theoretical description.
A pioneering contribution is due to Richmond \cite{Richmond1974,
Richmond1975}, who studied the effective interaction of two parallel planar
dielectric bodies with non-uniform surface charge distributions mediated by a
dilute electrolyte solution in between, \emph{assuming} that the linearized
Poisson-Boltzmann (Debye-H\"{u}ckel) approximation \cite{Debye1923} (see also
Refs.~\cite{Russel1989,McQuarrie2000,Hunter2001}) is applicable.
In recent years the issue of electrolyte solutions close to non-uniformly
charged substrates within the Debye-H\"{u}ckel approximation
\cite{BenYaakov2013,Ghosal2017,Mussotter2018,Sherwood2020},
(non-linearized) Poisson-Boltzmann theory \cite{Adar2016,Adar2018}, as well as
statistical field theory \cite{Naji2010,Naji2014,Ghodrat2015a,Ghodrat2015b,
Naji2018} has been addressed intensively (see also the review in
Ref.~\cite{Adar2017}).
These studies are focused on large length scales, either by ignoring the
microscopic fluid structure of the electrolyte solution or by modelling its
long-ranged structure within a square gradient approximation (see
Ref.~\cite{Mussotter2018}).
Moreover, microscopic approaches, e.g., Monte Carlo (MC) simulations
\cite{Bakhshandeh2015,Zhou2017} or classical density functional theory
(DFT) \cite{McCallum2017,Mussotter2020}, have been used.
But, due to technical reasons, theses studies were limited to rather small
systems and special types of surface charge non-uniformities.
Thus an approach is missing which exhibits the accuracy of a DFT combined
with the efficiency of a Debye-H\"{u}ckel approximation, in order to span the
whole range from microscopic to macroscopic length scales.

The present study suggests a step in this direction.
This novel method consists of a quadratic expansion of the density
functional not about the \emph{bulk} profiles (as within the
Debye-H\"{u}ckel approximation), but about the profiles of a
\emph{planar-symmetric} (i.e., quasi one-dimensional) system.
The surprising observation is, that microscopic hard-core contributions
turn out to be quantitatively irrelevant for the formation of the lateral
structure.
In this respect, disregarding the size of the fluid molecules and
ions by using the Debye-H\"{u}ckel approximation for laterally non-uniform
modes, as done in many previous studies, is justified.
However, the present investigation suggests, that, in contrast to those
previous studies, the Debye-H\"{u}ckel approximation should not be used for the
\emph{planar-symmetric} contributions, which require more
sophisticated descriptions including, e.g., finite size effects.

Our contribution is structured as follows: Section~\ref{sec:model}
describes the considered model of an electrolyte solution in contact with a
non-uniformly charged substrate and the formalism to infer the structural
quantities.
Results concerning the number density profiles in the normal and in the
lateral directions as well as concerning the interfacial tension
as function of the length scales of the lateral non-uniformities are
presented and discussed in Sec.~\ref{sec:results}.
Conclusions about the general structural features of electrolyte solutions in
contact with non-uniformly charged substrates are summarized in
Sec.~\ref{sec:conclusions}.


\section{\label{sec:model}Model and formalism}

\subsection{\label{sec:model:substrate}Non-uniformly charged substrate}

\begin{figure}
\includegraphics{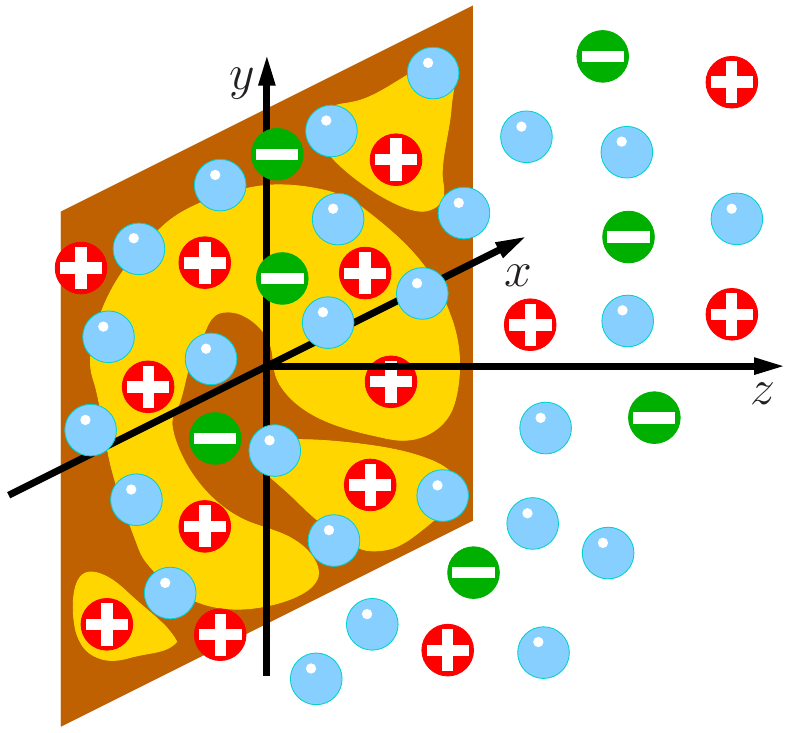}
\caption{A planar, non-uniformly charged substrate with negatively charged
(bright, yellow) and charge-neutral (dark, brown) regions is in
contact with a dilute univalent electrolyte solution (solvent blue, cations
red, anions green).
The $x$-$y$-plane of a three-dimensional Cartesian coordinate system coincides
with the substrate surface, whereas the $z$-direction points in normal
direction towards the bulk of the electrolyte solution.}
\label{fig:1}
\end{figure}

We consider a flat substrate with dielectric constant $\varepsilon_s$
the surface of which coincides with the $x$-$y$-plane of a
three-dimensional Cartesian coordinate system; the $z$-direction is
pointing towards the fluid at $z>0$ (see Fig.~\ref{fig:1}).
The substrate is non-uniformly charged with the surface charge density
$\sigma(\vec{u})$ at the lateral position $\vec{u}=(x,y)$.
In the present study periodic surface charge densities of the form
\begin{align}
   \sigma(\vec{u}) =
   \sum_{k,\ell\in\mathbb{Z}}
   \wh{\sigma}_{k\ell}\ \exp\left(\frac{2\pi i}{L}(k x + \ell y)
   \right),
   \label{eq:sigma}
\end{align}
are analyzed.
The Fourier coefficients $\wh{\sigma}_{k\ell}\in\mathbb{C}$, which fulfill
the constraints $\wh{\sigma}_{k,\ell}^* = \wh{\sigma}_{-k,-\ell}$ for
$\sigma(u)\in\mathbb{R}$, and the lateral length scale $L>0$ are free
parameters.
It will turn out that the periodicity of the lateral surface charge
distribution is of no physical relevance, but it is technically convenient.


\subsection{\label{sec:model:hardspheres}Charged hard spheres}

The charged substrate is in contact with a dilute univalent electrolyte
solution comprising three species of charged hard spheres: the solvent (species
$i=0$), cations (species $i=+$), and anions (species $i=-$).
Each species $i$ is characterized by its hard-core radius $R_i$ and the
valency $Z_i$ with $Z_0=0,Z_+=1,Z_-=-1$.
For simplicity, all radii are chosen to be equal, i.e., $R_0=R_+=R_-=:R$.
The bulk number densities of the electrolyte solution are given by
$\ol{\varrho}_0$ and $\ol{\varrho}_+=\ol{\varrho}_-=:I$, which is called
the ionic strength.
This leads to the packing fraction $\dps\eta = \frac{4\pi}{3}R^3(\ol{\varrho}_0
+ 2I)$.
From the Bjerrum length $\dps\ell_B := \frac{\beta e^2}{4\pi\varepsilon_0
\varepsilon_f}$, which is expressed in terms of the thermal energy
$\beta^{-1} = k_BT$, the elementary charge $e$, the vacuum electric
permittivity $\varepsilon_0$, and the fluid dielectric constant
$\varepsilon_f$, one obtains the Debye length $\kappa^{-1}$ with $\kappa^2 =
8\pi\ell_BI$.


\subsection{\label{sec:model:functional}Density functional method}

Close to the substrate the number density profile $\varrho_i(\vec{r})$ of
the fluid species $i$ varies as function of the position $\vec{r} =
(x,y,z) = (\vec{u},z)$, whereas $\varrho_i(\vec{u},z\to\infty)\to
\ol{\varrho}_i$.
The set of all three number density profiles is abbreviated by
$\varrho := (\rho_0,\rho_+,\rho_-)$.
The equilibrium number density profiles minimize the grand potential density
functional $\Omega[\varrho]$ \cite{Evans1979,Evans1990,Evans1992}, which, in
the present investigation, is approximated by
\begin{align}
   \beta\Omega[\varrho] = \int\d^3r\bigg\{
   &\
   \sum_i \varrho_i(\vec{r})\left[
   \ln\left(\frac{\varrho_i(\vec{r})}{\zeta_i}\right)-1+\beta V_i(z)\right]
   \notag\\
   &\
   +\Phi(n(\vec{r}))
   +\frac{\beta\varepsilon_0\varepsilon(z)}{2}
     \left(\nabla\psi(\vec{r})\right)^2\bigg\}.
   \label{eq:df1}
\end{align}
Here and in the following the common convention is in place that a
$d$-dimensional integration runs over $\mathbb{R}^d$ unless the
integration domain is specified.
Equation~\eqref{eq:df1} is to be understood as an asymptotic relation in the
thermodynamic limit, i.e., first all calculations are performed in a
finite domain which is extended to $\mathbb{R}^3$ subsequently.
The thermodynamic limit is guaranteed to exist, i.e., $\beta\Omega[\varrho]$
scales as the volume of the system, because the number density profiles
$\rho_i(\vec{r})$ are bounded due to the imposed lateral periodicity
of the system (see Eq.~\eqref{eq:sigma}) and due to the bulk limits
$\varrho_i(z\to\pm\infty)$.
In Eq.~\eqref{eq:df1}, $\dps\zeta_i=\Lambda_i^{-3}\exp(\beta\mu_i)$,
with the thermal wavelength $\Lambda_i$ and the chemical potential $\mu_i$,
denotes the (bulk) fugacity of species $i\in\{0,+,-\}$.
The hard-wall potential
\begin{align}
   V_i(z) = \begin{cases}
   \infty & \text{for $z \leq R_i$} \\
   0      & \text{for $z > R_i$}
   \end{cases}
   \label{eq:hardwall}
\end{align}
implies that the fluid particles cannot penetrate into the
substrate.
The hard-core interaction among the fluid particles is described in terms
of the White-Bear (mark I) excess free energy \cite{Roth2002}, which is given
by an excess free energy density $\Phi(n(r))$ expressed in terms of ten
weighted densities
\begin{align}
   n_\alpha(\vec{r})
   = \sum_i\int\d^3r'\,\omega_{\alpha,i}(\vec{r}-\vec{r'})
   \varrho_i(\vec{r'})
   \label{eq:nalpha}
\end{align}
that are indexed by $\alpha$ and that follow from the number density
profiles $\varrho_i$ via the weight functions $\omega_{\alpha,i}$. 
The electrostatic potential $\psi(\vec{r})$ fulfills Gau{\ss}'s law
\begin{align}
   \nabla\cdot\left(-\varepsilon_0\varepsilon(z)\nabla\psi(\vec{r})\right) =
   \sigma(\vec{u})\delta(z) + Q(\vec{r}),
   \label{eq:gauss}
\end{align}
where $Q(\vec{r}) := e\sum_i Z_i\varrho_i(\vec{r})$ and
\begin{align}
   \varepsilon(z) = \begin{cases}
   \varepsilon_s & \text{for $z\leq0$} \\
   \varepsilon_f & \text{for $z>0$}
   \end{cases}
   \label{eq:epsilonz}
\end{align}
with the boundary conditions
\begin{align}
   \frac{\p\psi}{\p z}\Big|_{(\vec{u},z=-\infty)} &= 0,
   \label{eq:bc1}\\
   -\varepsilon_0\left(
   \varepsilon_f\frac{\p\psi}{\p z}\Big|_{(\vec{u},z=0^+)} -
   \varepsilon_s\frac{\p\psi}{\p z}\Big|_{(\vec{u},z=0^-)}
   \right) &= \sigma(\vec{u}),
   \label{eq:bc2}\\
   \psi\Big|_{(\vec{u},z=\infty)} &= 0
   \label{eq:bc3}
\end{align}
for all $\vec{u}\in\mathbb{R}^2$ in lateral direction.
In order to guarantee the existence of the thermodynamic limit we consider
a globally charge-neutral system; actually Lebowitz and Lieb have shown that
slightly weaker but rather artificial conditions would also suffice
\cite{Lebowitz1969}.
Globally charge-neutral systems exhibit the gauge symmetry $\psi \mapsto \psi
+ \text{const}$, which is used to fix the value of the electrostatic potential
at $z=\infty$ by means of the Dirichlet boundary condition (see
Eq.~\eqref{eq:bc3}).
This implies the Neumann boundary condition $\dps\frac{\p\psi}{\p z}
\Big|_{(\vec{u},z=\infty)} = 0$.
For a globally charge-neutral system the electric displacement
$\dps-\varepsilon_0\varepsilon\frac{\p\psi}{\p z}$ has to be the same at
$z=-\infty$ and at $z=\infty$, which leads to Eq.~\eqref{eq:bc1}.
Finally, Eq.~\eqref{eq:bc2}, which is obtained by integrating
Eq.~\eqref{eq:gauss} over an infinitessimally small box around the point
$(\vec{u},z=0)$, describes the discontinuity of the electric
displacement at the charged surface $z=0$.

The equilibrium number density profiles $\varrho_i(\vec{r})$ vanish for
$z\leq R_i$ due to the hard wall (see Eq.~\eqref{eq:hardwall}),
whereas for $z > R_i$ they fulfill the Euler-Lagrange equations
\begin{align}
   0
   =&\
   \frac{\delta\beta\Omega}{\delta\rho_i(\vec{r})}[\varrho]
   \notag\\
   =&\
   \ln\left(\frac{\varrho_i(\vec{r})}{\zeta_i}\right)
   + \beta eZ_i\psi(\vec{r}) +
   \notag\\
   &\
   \sum_\alpha\int\d^3r'\,
   \frac{\p\Phi}{\p n_\alpha}(n(\vec{r'}))\,\omega_{\alpha,i}
   (\vec{r'}-\vec{r}).
   \label{eq:ele1}
\end{align}

The set of equations \eqref{eq:gauss} and \eqref{eq:bc1}--\eqref{eq:ele1} is
technically too demanding to be solvable numerically for an arbitrary
lateral length scale $L$.
In order to proceed, Eqs.~\eqref{eq:gauss} and \eqref{eq:bc1}--\eqref{eq:ele1}
are first solved for the laterally uniform charge distribution $\sigma^{(1)} :=
\wh{\sigma}_{00}$, which renders the quasi one-dimensional number density
profiles, the weighted densities, and the electrostatic potential denoted
as $\varrho_i^{(1)}(z)$, $n_\alpha^{(1)}(z)$, and $\psi^{(1)}(z)$,
respectively.

The quadratic expansion of the density functional in Eq.~\eqref{eq:df1} about
$\varrho_i^{(1)}(z)$ in terms of $\Delta\varrho_i(\vec{u},z) :=
\varrho_i(\vec{r}=(\vec{u},z))-\varrho_i^{(1)}(z)$ yields the approximation
$\Omega[\varrho] \approx\Omega[\varrho^{(1)}]+\Delta\Omega[\Delta\varrho]$ with
$\Delta\varrho := (\Delta\varrho_0,\Delta\varrho_+,\Delta\varrho_-)$, where
\begin{align}
   &\beta\Delta\Omega[\Delta\varrho]=
   \label{eq:df2}\\
   &\ \frac{1}{2}\int\d^3r\,\bigg(
   \sum_i\frac{(\Delta\varrho_i(\vec{r}))^2}{\varrho_i^{(1)}(z)}
   + \beta\varepsilon_0\varepsilon(z)\,(\nabla\Delta\psi(\vec{r}))^2 +
   \notag\\
   &\ \phantom{\frac{1}{2}\int\d^3r\,\bigg(}
   \sum_{\alpha,\alpha'}\frac{\p^2\Phi}{\p n_\alpha\p n_{\alpha'}}
   (n^{(1)}(z))\,\Delta n_\alpha(\vec{r})\Delta n_{\alpha'}(\vec{r})
   \bigg)
   \notag
\end{align}
with $\Delta n_\alpha(\vec{u},z) = n_\alpha(\vec{u},z)-n^{(1)}_\alpha(z)$ and
$\Delta\psi(\vec{u},z) = \psi(\vec{u},z)-\psi^{(1)}(z)$.
(Note that here ``$\Delta$'' is \emph{not} the Laplace operator
$\vec{\nabla}^2$).

The equilibrium profiles $\Delta\rho_i(\vec{r})$ fulfill the Euler-Lagrange
equations
\begin{align}
   0
   =&\
   \frac{\delta\beta\Delta\Omega}{\delta\Delta\rho_i(\vec{r})}[\Delta\varrho]
   \notag\\
   =&\
   \frac{\Delta\varrho_i(\vec{r})}{\rho^{(1)}_i(z)}
   + \beta eZ_i\,\Delta\psi(\vec{r}) +
   \label{eq:ele2}\\
   &\
   \sum_{\alpha,\alpha'}\int\d^3r'\,
   \frac{\p^2\Phi}{\p n_\alpha \p n_{\alpha'}}(n^{(1)}(z'))\,
   \omega_{\alpha,i}(\vec{r'}-\vec{r})\Delta n_{\alpha'}(\vec{r'})
   \notag
\end{align}
for $z > R_i$.

Introducing the lateral Fourier-transform
\begin{align}
   \wh{f}(\vec{q}) := \int\d^2u\,f(\vec{u})\exp(-i\vec{q}\cdot\vec{u}),
   \quad\vec{q}\in\mathbb{R}^2,
   \label{eq:fourier}
\end{align}
for functions $f(\vec{u})$ of the lateral coordinates $\vec{u}\in\mathbb{R}^2$,
from Eq.~\eqref{eq:ele2} one obtains 
\begin{align}
   0
   =&
   \frac{\Delta\wh{\varrho}_i(\vec{q},z)}{\rho^{(1)}_i(z)}
   + \beta eZ_i\,\Delta\wh{\psi}(\vec{q},z) +
   \label{eq:ele3}\\
   &
   \sum_{\alpha,\alpha'}\!\int\!\!\d z'\!
   \frac{\p^2\Phi}{\p n_\alpha \p n_{\alpha'}}(n^{(1)}(z'))\,
   \wh{\omega}_{\alpha,i}(\vec{q},z'\!-\!z)
   \Delta \wh{n}_{\alpha'}(\vec{q},z')
   \notag
\end{align}
for $z > R_i$ with
\begin{align}
   \Delta\wh{n}_\alpha(\vec{q},z)
   = \sum_i\int\d z'\,\wh{\omega}_{\alpha,i}(\vec{q},z-z)
   \Delta\wh{\varrho}_i(\vec{q},z').
   \label{eq:nalphahat}
\end{align}
Moreover, the lateral Fourier transformation of Gau{\ss}'s law
(see Eq.~\eqref{eq:gauss}) leads, due to Eq.~\eqref{eq:epsilonz},
to the Helmholtz equations
\begin{alignat}{1}
   \frac{\p^2\Delta\wh{\psi}}{\p z^2}(\vec{q},z)
   - |\vec{q}|^2\Delta\wh{\psi}(\vec{q},z)
   &=
   -\frac{\Delta\wh{Q}(\vec{q},z)}{\varepsilon_0\varepsilon_f}
   \notag\\
   &\qquad\text{for $z>0$,}
   \label{eq:helmholtzf}\\
   \frac{\p^2\Delta\wh{\psi}}{\p z^2}(\vec{q},z)
   - |\vec{q}|^2\Delta\wh{\psi}(\vec{q},z)
   &=
   -\frac{\Delta\wh{Q}(\vec{q},z)}{\varepsilon_0\varepsilon_s}
   =
   0
   \notag\\
   &\qquad\text{for $z<0$,}
   \label{eq:helmholtzs}   
\end{alignat}
where $\dps\Delta\wh{Q}(\vec{q},z)=e\sum_iZ_i\Delta\wh{\varrho}_i
(\vec{q},z)$.
Finally, the boundary conditions Eqs.~\eqref{eq:bc1}--\eqref{eq:bc3} take the
form
\begin{align}
   \frac{\p\Delta\wh{\psi}}{\p z}\Big|_{(\vec{q},z=-\infty)} &= 0,
   \label{eq:bc1hat}\\
   \varepsilon_f\frac{\p\Delta\wh{\psi}}{\p z}\Big|_{(\vec{q},z=0^+)} -
   \varepsilon_s\frac{\p\Delta\wh{\psi}}{\p z}\Big|_{(\vec{q},z=0^-)} &=
   -\frac{\Delta\wh{\sigma}(\vec{q})}{\varepsilon_0},
   \label{eq:bc2hat}\\
   \Delta\wh{\psi}\Big|_{(\vec{q},z=\infty)} &= 0.
   \label{eq:bc3hat}
\end{align}
Here $\Delta\wh{\sigma}(\vec{q})$ is the lateral Fourier transform of the
non-uniform contribution $\Delta\sigma(\vec{u}) := \sigma(\vec{u}) -
\sigma^{(1)}$ to the surface charge density.

The Helmholtz equation at $z<0$ (see Eq.~\eqref{eq:helmholtzs}) and the
Neumann boundary condition at $z=-\infty$ (see Eq.~\eqref{eq:bc1hat}) lead
to solutions of the form $\dps\Delta\wh{\psi}(\vec{q},z)=\Delta\wh{\psi}
(\vec{q},0)\exp(|\vec{q}|z)$ for $z<0$.
Then, from Eq.~\eqref{eq:bc2hat} one obtains the Robin boundary condition
\begin{align}
   \varepsilon_f\frac{\p\Delta\wh{\psi}}{\p z}\Big|_{(\vec{q},z=0^+)} -
   \varepsilon_s|\vec{q}|\Delta\wh{\psi}\Big|_{(\vec{q},z=0)} =
   -\frac{\Delta\wh{\sigma}(\vec{q})}{\varepsilon_0},
   \label{eq:bc4hat}
\end{align}
which, together with the Dirichlet boundary condition at $z=\infty$ (see
Eq.~\eqref{eq:bc3hat}), determines the solution of the Helmholtz equation
in Eq.~\eqref{eq:helmholtzf}.
Note that in the first term of Eq.~\eqref{eq:bc4hat} the upper limit of
$\dps\frac{\p\Delta\wh{\psi}}{\p z}$ occurs, because this quantity is
discontinuous at the surface $z=0$ due to Eq.~\eqref{eq:bc2hat}, whereas
in the second term of Eq.~\eqref{eq:bc4hat} $\Delta\wh{\psi}$ can be evaluated
at the surface, because the electrosatic potential is continuous
everywhere.

In the set of equations \eqref{eq:ele3}--\eqref{eq:bc4hat} the individual
Fourier modes, indicated by $\vec{q}$, are decoupled, and the remaining
$z$-coordinate normal to the substrate leads to a quasi one-dimensional
problem, which can be efficiently solved numerically.

Moreover, any function $f(\vec{u})$ with $f(x+L,y)=f(x,y+L)=f(x,y)$
for all $\vec{u}=(x,y)\in\mathbb{R}^2$ can be written as
\begin{align}
   f(\vec{u}=(x,y))
   =
   \sum_{k,\ell\in\mathbb{Z}}
   f_{k\ell}\exp\left(\frac{2\pi i}{L}(kx+\ell y)\right)
   \label{eq:fuper}
\end{align}
with the the Fourier transform
\begin{align}
   &\ \wh{f}(\vec{q}=(q_x,q_y)) =
   \notag\\
   &\ \sum_{k,\ell\in\mathbb{Z}}
   (2\pi)^2f_{k\ell}\,
   \delta\left(q_x-\frac{2\pi k}{L}\right)
   \delta\left(q_y-\frac{2\pi \ell}{L}\right),
   \label{eq:fhatqper}
\end{align}
which can be non-zero only for lateral wave numbers $\dps\vec{q} =
\vec{q}_{k\ell} := \frac{2\pi}{L}(k, l)$ with $k,\ell\in\mathbb{Z}$.
Therefore, the determination of the (approximate) equilibrium number
density profiles $\varrho_i(\vec{r})$ merely requires to calculate the
Fourier transforms $\Delta\wh{\varrho}_i(\vec{q},z)$ as solutions of
Eqs.~\eqref{eq:ele3}--\eqref{eq:bc4hat} for $\vec{q}=\vec{q}_{k\ell}$ with
$k,\ell\in\mathbb{Z}$.


\subsection{\label{sec:methods:interfacialtension}Interfacial tension}

Besides the profiles $\varrho_i(\vec{r})=\varrho_i^{(1)}(z)+\Delta
\varrho_i(\vec{r})$, $Q(\vec{r})=Q^{(1)}(z)+\Delta Q(\vec{r})$, and
$\psi(\vec{r})=\psi^{(1)}(z)+\Delta\psi(\vec{r})$, from
Eqs.~\eqref{eq:ele3}--\eqref{eq:bc4hat} the following discussion also addresses
the interfacial tension $\gamma$ as a common surface quantity.
Here it is defined w.r.t.\ the geometrical substrate surface at $z=0$.
If $\gamma^{(1)}$ is the interfacial tension of a uniformly charged substrate
with surface charge density $\sigma^{(1)}$, one obtains the deviation
$\Delta\gamma := \gamma - \gamma^{(1)}$ due to non-uniformities within the
quadratic approximation (see Eq.~\eqref{eq:df2}) as
\begin{align}
   \Delta\gamma =
   \frac{\Delta\Omega_L[\Delta\varrho]}{L^2} =
   \frac{1}{2L^2}\int\limits_{[0,L)^2}\d^2u\,\Delta\sigma(\vec{u})
   \Delta\psi(\vec{u},0),
   \label{eq:deltagamma}
\end{align}
where $\Delta\Omega_L$ means integration over $\vec{r}=(\vec{u},z)\in
[0,L)^2\times\mathbb{R}$ in Eq.~\eqref{eq:df2}, i.e., over one lateral periodic
image.
This expression can be obtained by multiplying Eq.~\eqref{eq:ele2} with
$\Delta\varrho_i(\vec{r})$, summing over $i$, integrating w.r.t.\ $\vec{r}$,
and inserting the resulting equation into Eq.~\eqref{eq:df2}.


\subsection{\label{sec:methods:parameters}Parameters}

The main focus of the present study is the dependence of the profiles
$\varrho_i(\vec{r})$, $Q(\vec{r})$, and $\psi(\vec{r})$ as well as of the
interfacial tension $\gamma$ on the characteristic length scale $L$ of the
lateral charge non-uniformities.
The remaining numerous model parameters are fixed to certain realistic
values.

\begin{figure}
\includegraphics{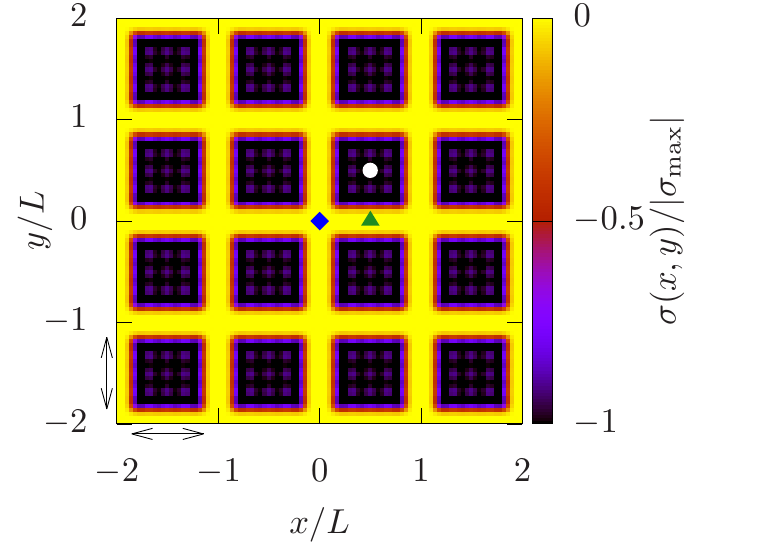}
\caption{The non-uniform surface charge density $\sigma(x,y)$ comprising
25 Fourier modes (see Eq.~\eqref{eq:sigmafinite}) considered in the present
investigation is a continuous approximation of a substrate with half of its
area being charge-neutral and the other half being made up of charged square
patches of side length $L/\sqrt{2}$ (indicated by the double arrows in the
lower left corner) with surface charge density $\sigma_\text{max}$.
The restriction to a finite number of Fourier modes gives rise to slight
artifacts such as smooth instead of step-like variations as well as undulations
(see the apparent substructure in the dark square areas) instead of
plateaus.
The mean surface charge density is $\dps\sigma^{(1)} =
\frac{\sigma_\text{max}}{2}$.
The charged patches are arranged on a (two-dimensional) square lattice with
periodicity $L$, which sets the lateral length scale of this structure.
Figure~\ref{fig:3} displays the number density profiles $\varrho_i(\vec{u},
z)$ along the $z$-direction at three lateral positions $\vec{u}=(x,y)$:
$\vec{u}=(0,0)$ (blue diamond), $\vec{u}=(L/2,0)$ (green triangle),
and $\vec{u}=(L/2,L/2)$ (white dot).}
\label{fig:2}
\end{figure}

As a non-trivial surface structure we choose a two-dimensional square
lattice with periodicity $L>0$ such that the surface charge density takes the
constant value $\sigma_\text{max}$ for one half of the surface and $0$ for the
other half.
This leads to an average surface charge density $\dps\sigma^{(1)} =
\frac{\sigma_\text{max}}{2}$ and in Eq.~\eqref{eq:sigma} to the Fourier
coefficients 
\begin{align}
   \wh{\sigma}_{k\ell} =
   \sigma^{(1)}\,(-1)^{k+\ell}
   \,\sinc\left(\frac{\pi k}{\sqrt{2}}\right)
   \,\sinc\left(\frac{\pi \ell}{\sqrt{2}}\right),
   \label{eq:sigmahat}
\end{align}
where the $\sinc$ function is defined as $\dps\sinc(t)=\frac{\sin(t)}{t}$
for $t\not=0$ and $\sinc(t)=1$ for $t=0$.
However, in order to limit the computational demand only Fourier modes
with $|k|,|\ell| \leq 5$ are used here.
The resulting surface charge density
\begin{align}
   \sigma(\vec{u}) :=
   \sum_{k,\ell=-5}^5\wh{\sigma}_{k\ell}
   \exp\left(\frac{2\pi i}{L}(kx+\ell y)\right)
   \label{eq:sigmafinite}
\end{align}
is a continuous approximation of the actually considered step-like
structure (see Fig.~\ref{fig:2}).

In addition to the thermal energy $\beta^{-1}$ as the energy unit and the
elementary charge $e$ as the charge unit, the Debye length $\kappa^{-1}$
is chosen as the length unit.
Setting the fluid particle radii to be equal, i.e., $R_0 = R_+ = R_- = R$,
the model comprises the following six dimensionless parameters:
\begin{align}
   &\ L^*:=\kappa L,\
   \sigma^*:=\frac{\sigma^{(1)}}{e\kappa^2},\
   \eta=\frac{4\pi}{3}R^3(\ol{\varrho}_0+\ol{\varrho}_++\ol{\varrho}_-),
   \notag\\
   &\ \kappa R,\
   \kappa\ell_B,\ \text{and}\
   \frac{\varepsilon_s}{\varepsilon_f}.
   \label{eq:parameters}
\end{align}
In the following, the dependence of structural quantities on $L^*$
over two decades is discussed, and two values of the parameter
$\dps\sigma^*\in\{-1.1,-3.3\}$ are considered.

For the remaining parameters in Eq.~\eqref{eq:parameters}, fixed values are
chosen according to an aqueous solution $(\ol{\varrho}_0\approx56\,\m{M}, R
\approx0.13\,\m{nm}, \ell_B\approx0.7\,\m{nm}, \varepsilon_f\approx80)$ with
ionic strength $I \approx 8.5\,\m{mM}$, i.e., $\kappa\approx0.3\,\m{nm}^{-1}$,
in contact with a substrate with dielectric constant $\epsilon_s\approx8$:
\begin{align}
   \eta \approx 0.3, \
   \kappa R \approx 0.039, \
   \kappa \ell_B\approx 0.21, \ \text{and}\
   \frac{\varepsilon_s}{\varepsilon_f} \approx 0.1.
   \label{eq:values}
\end{align}
Note that here number densities are specified as molar concentrations in
moles per liter: $1\,\m{M} = 1\,\m{mol\,dm^{-3}} \approx 0.6022\,\m{nm}^{-3}$.

Given an aqueous electrolyte solution in contact with a uniformly charged
surface the saturation surface charge density $\dps \sigma_\text{sat} =
\frac{e\kappa}{\pi\ell_B}$ denotes the crossover between a weakly charged
surface with $|\sigma^{(1)}|<\sigma_\text{sat}$, for which the linearized
Poisson-Boltzmann (i.e., Debye-H\"{u}ckel) equation is applicable, and a
strongly charged surface with $|\sigma^{(1)}| > \sigma_\text{sat}$, for which
the full non-linear Poisson-Boltzmann equation is required \cite{Bocquet2002}.
For the aqueous electrolyte solution specified above, the saturation
surface charge density is given by $\sigma_\text{sat}\approx 2.2\,\mu\m{C}\,
\m{cm}^{-2}$, which corresponds to a crossover value $\dps\sigma^*_\text{sat}
:= \frac{\sigma_\text{sat}}{e\kappa^2}=\frac{1}{\pi\kappa\ell_B}\approx1.5$.
The two values $\sigma^*=-1.1$ and $-3.3$, which will be considered in the
following, have been chosen to represent the cases of weakly and strongly
charged surfaces, respectively.


\section{\label{sec:results}Results and Discussion}

\subsection{\label{sec:results:normal}Normal profiles}

\begin{figure*}
\includegraphics{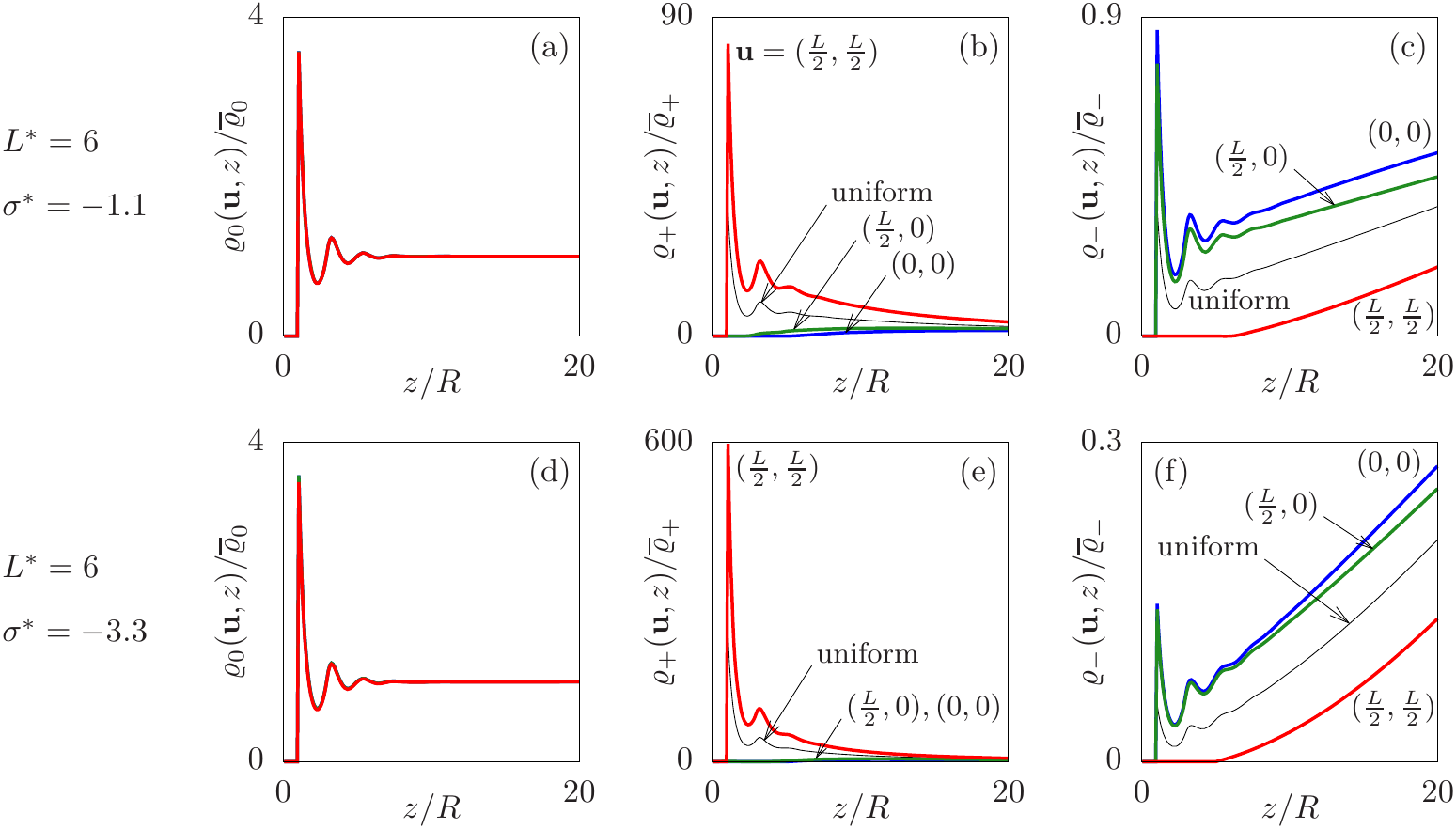}
\caption{The plots show the number density profiles $\varrho_i(\vec{u},z)$
of the solvent ($i=0$) (see the panels (a) and (d)), of the cations
($i=+$) (see the panels (b) and (e)), and of the anions ($i=-$) (see
the panels (c) and (f)) close to a planar, non-uniformly charged substrate
corresponding to Fig.~\ref{fig:2} with periodicity $L^* = \kappa L = 6$
and mean surface charge $\dps\sigma^* = \frac{\sigma^{(1)}}{e\kappa^2} =
-1.1$ (see the panels (a)--(c)) and $-3.3$ (see the panels (d)--(f)).
The number density profiles $\varrho_i(\vec{u},z)$ are given as functions of
the normal coordinate $z>0$ for three representative lateral positions
in the lateral elementary cell $[0,L)\times[L,0)$ (see Fig.~\ref{fig:2}): at
the origin $\vec{u}=(0,0)$ (blue curves corresponding to the blue
diamond), close to an edge $\vec{u}=(L/2,0)$ (green curves
corresponding to the green triangle), and at the center point $(L/2,L/2)$
(red curves corresponding to the white dot).
For comparison the number density profiles $\varrho_i^{(1)}(z)$ for a uniform
charge distribution with surface charge density $\sigma^{(1)}$ is shown
(thin black curves).
For $z\to\infty$ all profiles approach the corresponding bulk number densities
$\ol{\varrho}_i$.
Close to the substrate the typical layering due to the hard cores of the fluid
particles is clearly visible.
Whereas the solvent number density $\varrho_0(\vec{u},z)$ (see panels (a) and
(d)) varies barely as function of the lateral position $\vec{u}$
or the surface charge $\sigma^*$, the cation number density
$\varrho_+(\vec{u},z)$ (see panels (b) and (e)) and the anion number densities
$\varrho_-(\vec{u},z)$ (see panels (c) and (f)) are sensitive to both $\vec{u}$
and $\sigma^*$.
Note that the surface charge is negative here, i.e., $\sigma^*<0$, so that the
cations accumulate at and the anions are depleted from lateral positions close
to the charged square patches of the substrate (see Fig.~\ref{fig:2}).}
\label{fig:3}
\end{figure*}

Figure~\ref{fig:3} displays the number density profiles $\rho_i(\vec{u},z),i\in
\{0,+,-\}$, as functions of the normal coordinate $z>0$ for three
characteristic lateral positions $\vec{u}=(x,y)=(0,0)$ (blue curves, blue
diamond in Fig.~\ref{fig:2}), $(L/2,0)$ (green curves, green triangle
in Fig.~\ref{fig:2}), and $(L/2,L/2)$ (red curves, white dot in
Fig.~\ref{fig:2}) at a corner, at an edge, and at the center of the lateral
elementary cell $[0,L)\times[0,L) \subseteq \mathbb{R}^2$, respectively, for
$L^*=6$.
Panels (a)--(c) show the case $\sigma^*=-1.1$ whereas panels (d)--(f) show the
case $\sigma^*=-3.3$.
For comparison the corresponding profiles $\varrho_i^{(1)}(z)$ close to a
uniformly charged substrate are depicted (see the thin black curves).
It can be observed that the solvent number density profiles $\varrho_0(\vec{u},
z)$ (see Figs.~\ref{fig:3}(a) and (d)) are largely insensitive to the
lateral position $\vec{u}$ and to the magnitude of the surface charge density
$|\sigma^*|$, because the solvent particles in the present model are
electrically neutral and non-polar.
Within a model for a polar solvent one can expect variations of $\varrho_0
(\vec{u},z)$ to occur upon changing $\vec{u}$ or $\sigma^*$.

As the surface charge is negative, the cation number densities
$\varrho_+(\vec{u},z)$ close to the substrate surface are larger than in the
bulk, whereas the anion number density profiles $\varrho_-(\vec{u},z)$ close to
the substrate are smaller than in the bulk.
As expected, these trends are particularly pronounced for highly charged
surfaces, i.e., large values of $|\sigma^*|$, and at lateral positions
$\vec{u}$ corresponding to highly charged regions on the substrate.

According to Eq.~\eqref{eq:ele3} the lateral structure, expressed in terms of
$\Delta\wh{\varrho}_i(\vec{q},z)$, is determined by the electrostatic
potential, represented by $\Delta\wh{\psi}(\vec{q},z)$, as well as by the
hard-core interaction, given by the third expression in
Eq.~\eqref{eq:ele3}.
Upon ignoring the hard-core contribution one obtains approximate lateral number
density variations
\begin{align}
   \Delta\wh{\varrho}_i(\vec{q},z)
   \approx
   \Delta\wh{\varrho}_i^\text{DH}(\vec{q},z)
   :=
   -\beta e Z_i\varrho_i^{(1)}(z)\Delta\wh{\psi}(\vec{q},z),
   \label{eq:dh1}
\end{align}
which resemble those within linear Poisson-Boltzmann (i.e., Debye-H\"{u}ckel)
theory.
Inverse Fourier transformation leads to
\begin{align}
   \Delta\varrho_i^\text{DH}(\vec{u},z)
   =
   -\beta e Z_i\varrho_i^{(1)}(z)\Delta\psi(\vec{u},z)
   \label{eq:dh2}
\end{align}
so that
\begin{align}
   \varrho_i^\text{DH}(\vec{u},z)
   :=&\ \varrho_i^{(1)}(z) + \Delta\varrho_i^\text{DH}(\vec{u},z)
   \notag\\
   =&\ \varrho_i^{(1)}(z)\left(1 - \beta e Z_i \Delta\psi(\vec{u},z)\right).
   \label{eq:dh3}
\end{align}

\begin{figure*}
\includegraphics{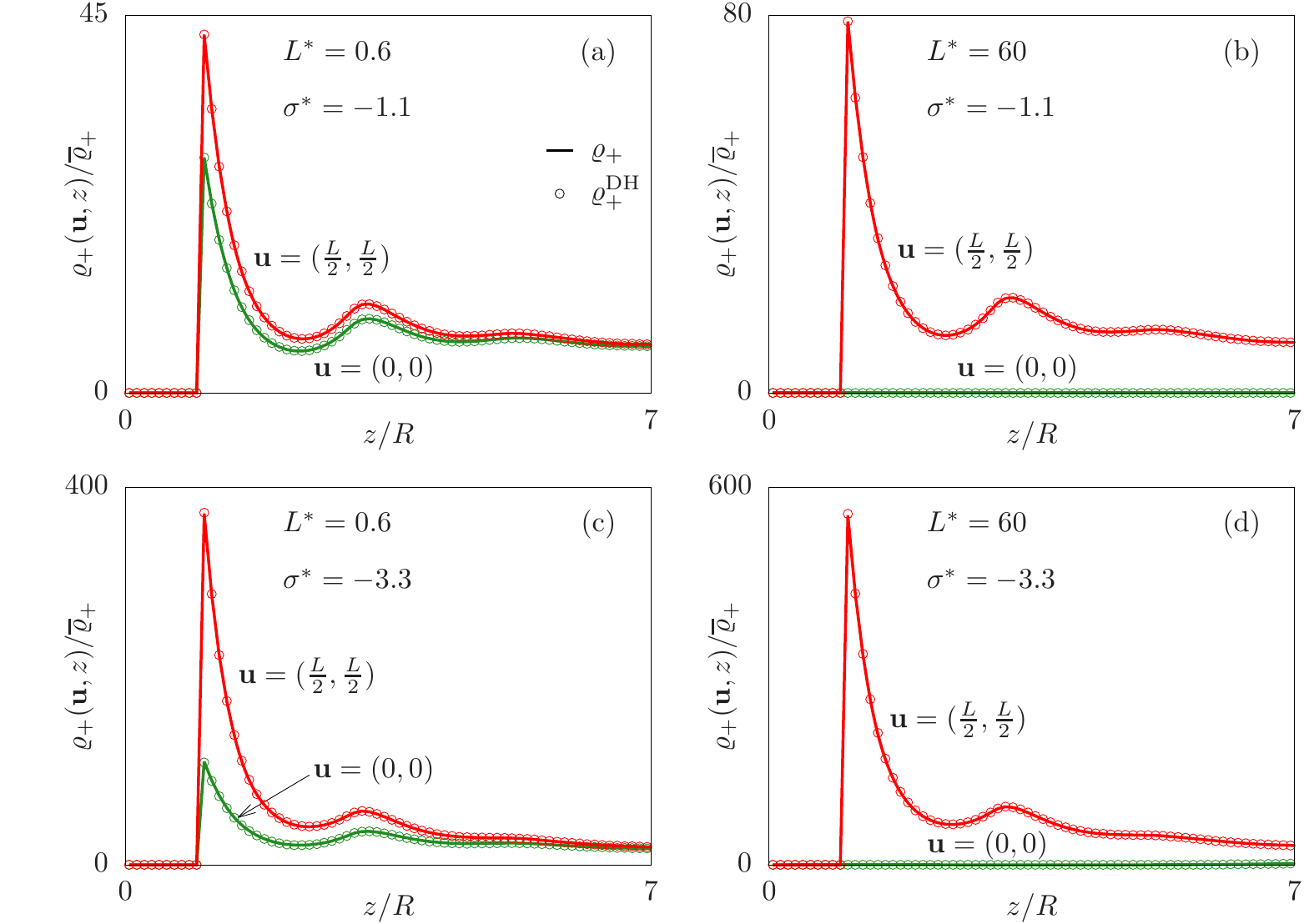}
\caption{In order to assess the approximation $\varrho_i(\vec{r})\approx
\varrho_i^\text{DH}(\vec{r})$ for the full number density profiles
$\varrho_i(\vec{r})$ (solid curves) by the Debye-H\"{u}ckel profiles
$\varrho_i^\text{DH}(\vec{r})$ (circles) as defined in Eq.~\eqref{eq:dh3},
the cation profiles ($i=+$) are shown for the two values $L^*=0.6$ (see
panels (a) and (c)) and $L^*=60$ (see panels (b) and (d)), the two values
$\sigma^*=-1.1$
(see panels (a) and (b)) and $\sigma^*=-3.3$ (see panels (c) and (d)), as
well as for two lateral positions $\vec{u}=(0,0)$ (green curves) and
$(L/2,L/2)$ (red curves).
We find excellent quantitative agreement.
For the large length $L^*=60$ the number densities $\varrho_+(\vec{u},z)$
are close to the bulk number density $\ol{\varrho}_+$ at the lateral
position $\vec{u}=(0,0)$, where the substrate is uncharged within a radius of
a few Debye lengths (see the green curves and circles in panels (b)
and (d)).}
\label{fig:4}
\end{figure*}

Figure~\ref{fig:4} compares the full number density profiles
$\varrho_i(\vec{u},z)$ (solid curves) with the corresponding Debye-H\"{u}ckel
approximations $\varrho_i^\text{DH}(\vec{u},z)$ (circles) according to
Eq.~\eqref{eq:dh3} at the lateral positions $\vec{u}=(0,0)$, i.e., at
the origin (in green), and at $\vec{u}=(L/2,L/2)$, i.e., in the center of
the elementary cell (in red), for lateral length scales $L^*\in\{0.6,
60\}$, and surface charges $\sigma^*\in\{-1.1, -3.3\}$.
It turns out, that the approximation $\varrho_i(\vec{r})\approx
\varrho_i^\text{DH}(\vec{r})$ is reliable to a high degree, i.e., the
hard-core contribution as the last term in Eq.~\eqref{eq:ele3} can be
safely ignored.
Whereas the hard-core interaction plays an important role for the number
density profiles $\varrho_i^{(1)}(z)$ close to laterally uniformly charged
substrates, it does not influence the lateral structure formation
significantly.

Since the hard-core contribution as the last term in
Eq.~\eqref{eq:ele3} is quantitatively negligible, one ends up with the
approximation
\begin{align}
   \Delta\wh{Q}(\vec{q},z)
   \approx&\
   e\sum_iZ_i\left(-\beta eZ_i\varrho^{(1)}_i(z)\right)
   \Delta\wh{\psi}(\vec{q},z)
   \notag\\
   =&\
   -\beta e^2\sum_i Z_i^2\varrho^{(1)}_i(z)
   \Delta\wh{\psi}(\vec{q},z)
   \notag\\
   =&\
   -\beta e^2\left(\varrho^{(1)}_+(z)+\varrho^{(1)}_-(z)\right)
   \Delta\wh{\psi}(\vec{q},z)
   \label{eq:Qhat}
\end{align}
which is equally valid.

\begin{figure}
\includegraphics{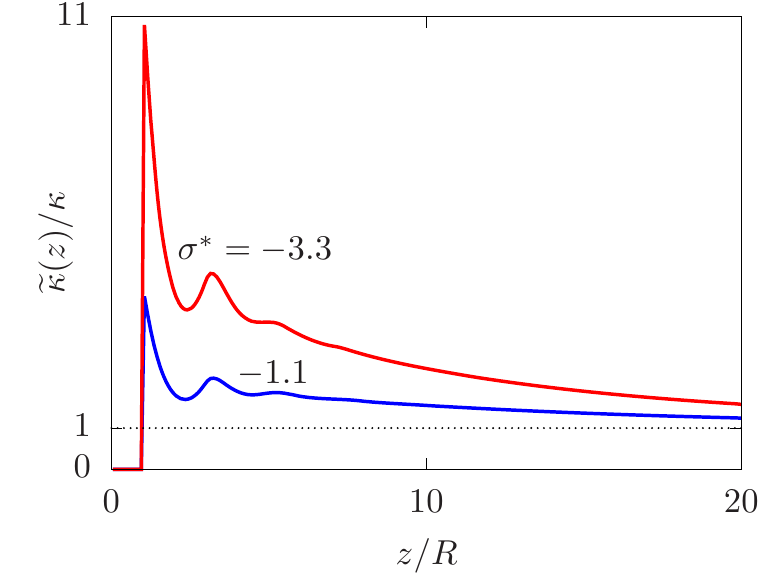}
\caption{For $z\to\infty$ the function $\wt{\kappa}(z)$, defined in
Eq.~\eqref{eq:kappatilde}, approaches the inverse Debye length
$\kappa$.
Beyond the hard-core layering range, $\wt{\kappa}(z)/\kappa$ attains unity
within a few particle radii $R$.}
\label{fig:5}
\end{figure}

Upon inserting Eq.~\eqref{eq:Qhat} into the Helmholtz equation for $z>0$
(see Eq.~\eqref{eq:helmholtzf}) one obtains
\begin{align}
   \frac{\p^2\Delta\wh{\psi}}{\p z^2}(\vec{q},z)
   = \left(|\vec{q}|^2 + \wt{\kappa}(z)^2\right)\Delta\wh{\psi}(\vec{q},z)
   \label{eq:helmholtzf2}
\end{align}
with the abbreviation
\begin{align}
   \wt{\kappa}(z) :=
   \sqrt{4\pi\ell_B\left(\varrho^{(1)}_+(z)+\varrho^{(1)}_-(z)\right)}
   \label{eq:kappatilde}.
\end{align}
For $z\to\infty$ the quantity $\wt{\kappa}(z)$ approaches the inverse
Debye length, i.e., $\wt{\kappa}(z)\to\kappa$, as in this limit
$\varrho_\pm^{(1)}(z) \to I$.
Figure~\ref{fig:5} shows, that $\wt{\kappa}(z)/\kappa$ attains its bulk
value $1$ already a few particle radii $R$ away from the substrate.

Hence beyond a few particle radii $R$ away from the substrate, i.e., at
$z \gg R$, Eq.~\eqref{eq:helmholtzf2} reduces to
\begin{align}
   \frac{\p^2\Delta\wh{\psi}}{\p z^2}(\vec{q},z)
   \simeq \left(|\vec{q}|^2 + \kappa^2\right)\Delta\wh{\psi}(\vec{q},z),
   \label{eq:helmholtzf3}
\end{align}
with the solution
\begin{align}
   \Delta\wh{\psi}(\vec{q},z) \propto
   \exp\left(-\frac{z}{\lambda(|\vec{q}|)}\right),
   \qquad\text{$z \gg R$,}
   \label{eq:psihat}
\end{align}
with the \emph{normal decay length}
\begin{align}
   \lambda(q) := \frac{1}{\sqrt{q^2 + \kappa^2}}.
   \label{eq:lambda}
\end{align}
According to Eq.~\eqref{eq:dh1}, in the range $z \gg R$ the modes of the
lateral structure $\Delta\wh{\varrho}_i(\vec{q},z)$ decay on the same normal
length scale $\lambda(|\vec{q}|)$.
Whereas the decay length $\lambda(q)$ is a bulk quantity, the
proportionality prefactor of the asymptotics in Eq.~\eqref{eq:psihat}
dependson the surface charge density (see Eq.~\eqref{eq:bc4hat}) as well as on
details of the ion number density profiles $\varrho^{(1)}_+(z)$ and
$\varrho^{(1)}_-(z)$ (see Eqs.~\eqref{eq:helmholtzf2} and
\eqref{eq:kappatilde}).

\subsection{\label{sec:results:lateral}Lateral profiles}

The length scale, on which the lateral modes $\vec{q}$ decay in the normal
direction, is given by $\lambda(|\vec{q}|)$ (see Eq.~\eqref{eq:lambda}).
It attains its maximum value $\kappa^{-1}$, i.e., the Debye length, at
$\vec{q}=0$.
Accordingly, the normal decay length $\lambda(|\vec{q}|)$ is not larger than
the Debye length $\kappa^{-1}$.
Upon increasing $|\vec{q}|$ the normal decay length $\lambda(|\vec{q}|)$
decreases monotonically.

\begin{figure}
\includegraphics{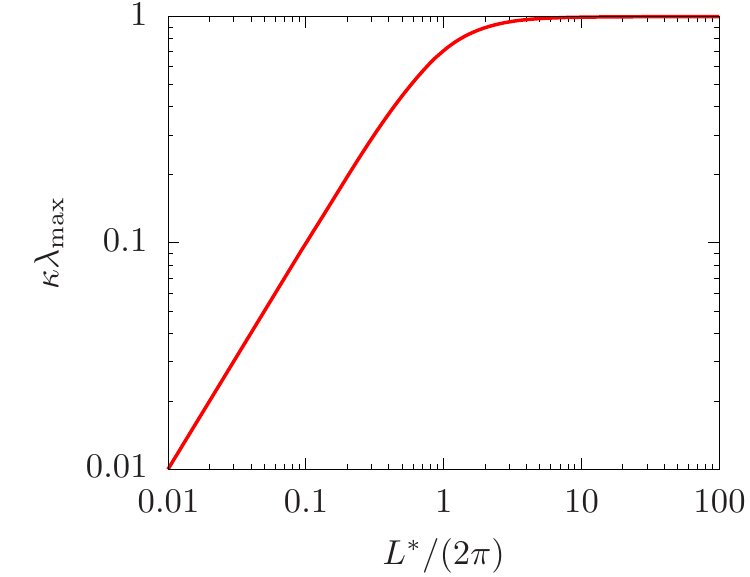}
\caption{At $L^*=\kappa L\approx 2\pi$ the largest normal decay length
$\lambda_\text{max}$ of the laterally non-uniform modes (see
Eq.~\eqref{eq:lambdamax}) crosses over from a linear regime at short length
scales $L^*\ll2\pi$ to $\lambda_\text{max}\simeq\kappa^{-1}$ (i.e., the
Debye length) at large length scales $L^*\gg2\pi$.}
\label{fig:6}
\end{figure}

Since $\sigma^{(1)} = \wh{\sigma}_{00}$ and
\begin{align}
   \Delta\sigma(\vec{u}=(x,y)) =
   \sigma(\vec{u})-\sigma^{(1)} =
   \notag\\
   \sum_{\begin{smallmatrix} k,\ell\in\mathbb{Z} \\
                             (k,\ell)\not=(0,0)\end{smallmatrix}}
   \wh{\sigma}_{k\ell}\ \exp\left(\frac{2\pi i}{L}(kx+\ell y)\right),
   \label{eq:deltasigmau}
\end{align}
i.e., $\Delta\wh{\sigma}(\vec{q}=\vec{q}_{00}=0) = 0$ due to
Eq.~\eqref{eq:fhatqper}, the smallest wave number $\dps |\vec{q}|
= |\vec{q}_{k\ell}| = \frac{2\pi}{L}\sqrt{k^2+\ell^2}$ contributing to a
lateral structure is $\dps q_\text{min} = |\vec{q}_{\pm1,0}|=|\vec{q}_{0,\pm1}|
=\frac{2\pi}{L}$.
Hence the lateral structure induced by a non-uniformly charged substrate
decays in normal direction on the length scale
\begin{align}
   \lambda_\text{max}
   &= \lambda(q_\text{min})
   = \frac{1}{\dps\sqrt{\left(\frac{2\pi}{L}\right)^2+\kappa^2}}
   = \frac{L}{\sqrt{(2\pi)^2 + (L^*)^2}}
   \notag\\
   &\simeq\begin{cases}
      \dps\frac{L}{2\pi}, & \text{for $L^*\ll2\pi$} \\[7pt]
      \kappa^{-1},        & \text{for $L^*\gg2\pi$.}
   \end{cases}
   \label{eq:lambdamax}
\end{align}
Figure~\ref{fig:6} displays the dependence of $\lambda_\text{max}$ on the
length scale parameter $L^*=\kappa L$.
At short length scales $L^*\ll2\pi$ a linear dependence is found, which crosses
over to $\lambda_\text{max}\simeq\kappa^{-1}$ (Debye length) at large length
scales $L^*\gg 2\pi$.

\begin{figure*}
\includegraphics{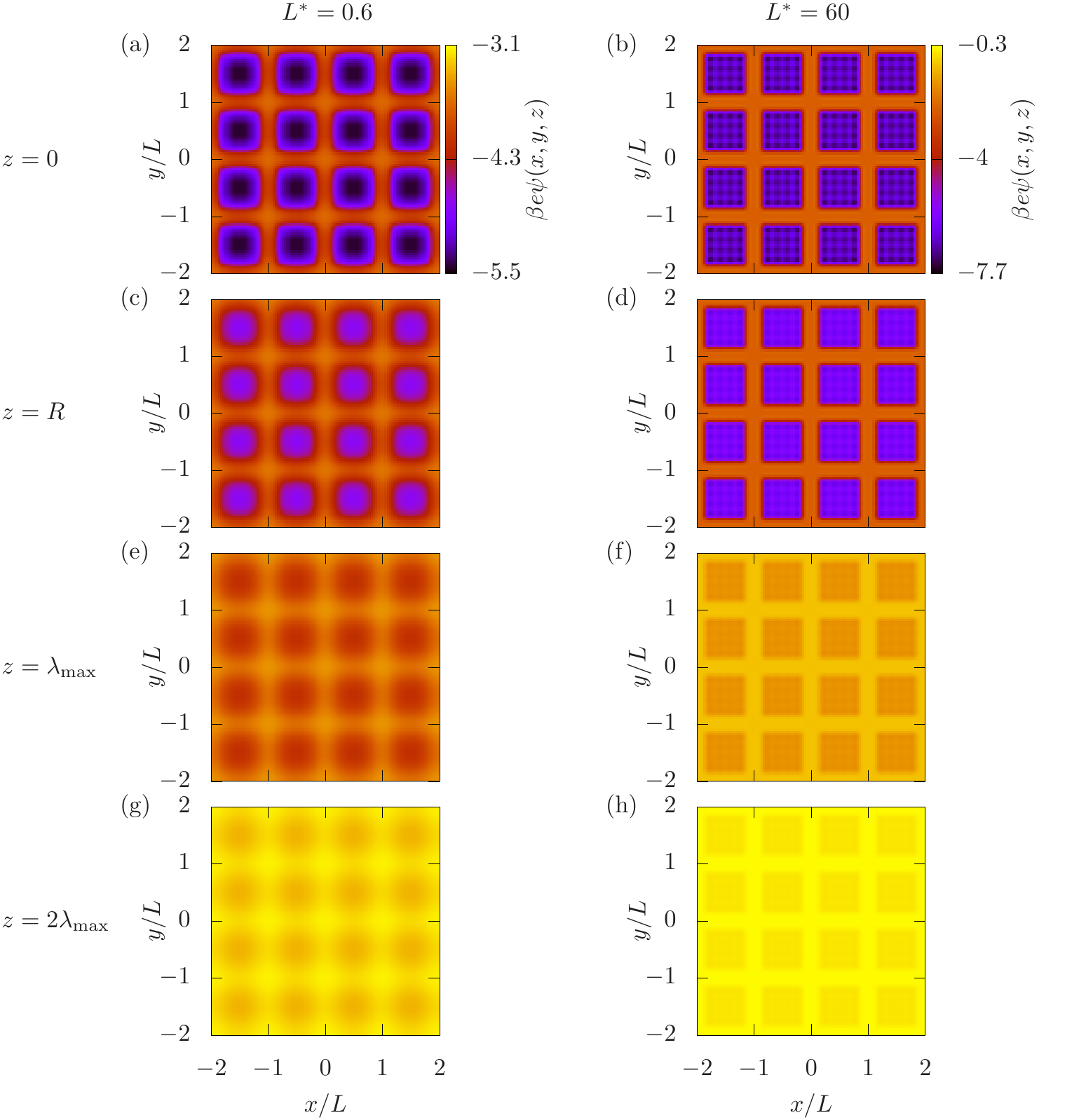}
\caption{The reduced electrostatic potential $\beta e\psi(\vec{u},z)$
as function of the lateral position $\vec{u}$ for fixed distances $z\geq0$ from
the substrate determines, and hence represents, the lateral structure of the
electrolyte solution (see Eq.~\eqref{eq:dh3}).
The left column corresponds to the case $L^*=0.6$ (short lateral length
scale) and the right column to the case $L^*=60$ (large lateral length
scale).
For each case, the lateral structure decays in normal direction on the length
scale of $\lambda_\text{max}$ (see Eq.~\eqref{eq:lambdamax}).
The contrast between charged and neutral parts of the substrate at $z=0$
is clearly visible (see panels (a) and (b)).
In the first contact layer of the fluid at $z=R$ (see panels (c) and (d)),
the contrast is still present, but slightly blurred.
At the distance $z=\lambda_\text{max}$ (see panels (e) and (f)) the contrast is
diminished substantially and even more so at $z=2\lambda_\text{max}$ (see
panels (g) and (h)).
The decay of the lateral structure as function of $z/\lambda_\text{max}$ is
similar, irrespective of the lateral length scale $L^*$.
Upon increasing $z$, for $L^*=0.6$ the rectangular shape of the charged
pattern is washed out in favor of circular patterns.
For $L^*=60$ the rectangular shape of the patterns remains even for $z=2
\lambda_\text{max}$.}
\label{fig:7}
\end{figure*}

From the quantitatively reliable approximation $\varrho_i(\vec{r})
\approx\varrho^\text{DH}_i(\vec{r})$ (see the previous
Subsec.~\ref{sec:results:normal} and in particular Fig.~\ref{fig:4}) one can
infer that the lateral structure of the electrolyte solution, i.e., 
$\varrho_i(\vec{r})$, is determined by the lateral structure of the
electrostatic potential $\psi(\vec{r})$ (see Eq.~\eqref{eq:dh3}).
Accordingly, Fig.~\ref{fig:7} displays two sequences of lateral profiles of the
electrostatic potential $\psi(\vec{u},z)$, i.e., functions of $\vec{u}$ with
$z$ fixed, at the normal positions $z = 0$, $R$,
$\lambda_\text{max}$, and $2\lambda_\text{max}$ for $L^*=0.6$ (left column:
(a), (c), (e), (g)) and $60$ (right column: (b), (d), (f), (h)).
Qualitatively, the difference of the electrostatic potential between
lateral positions associated with large and with small surface charge densities
diminishes with increasing distance from the substrate.
However, although the normal decay length $\lambda_\text{max}$ is very
different for the two cases ($\kappa\lambda_\text{max}\approx0.1$ for
$L^*=0.6$ and $\kappa\lambda_\text{max}\approx1$ for $L^*=60$), the decay of
the lateral structure of the two as function of $z/\lambda_\text{max}$ is
similar.

\subsection{\label{sec:results:tension}Interfacial tension}

\begin{figure}
\includegraphics{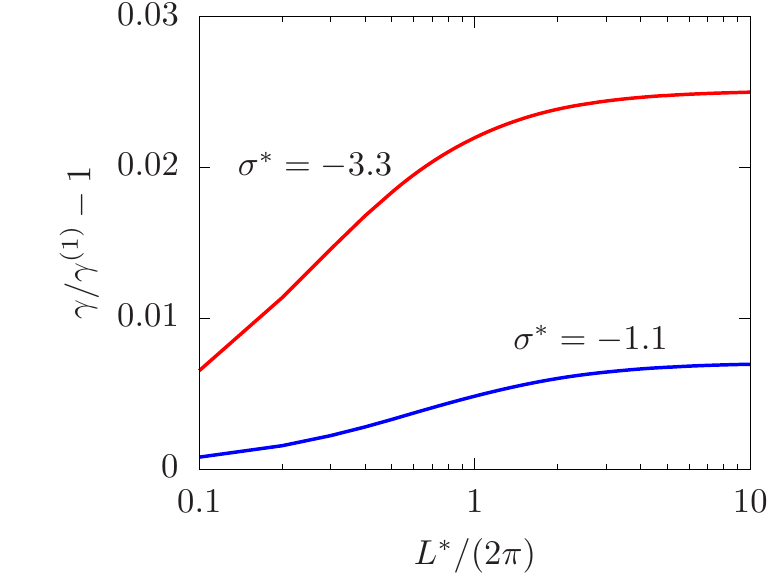}
\caption{Upon increasing the lateral length scale $L^*$ the interfacial tension
$\gamma$ increases with respect to its value $\gamma^{(1)}$ of a uniformly
charged substrate.
In sync with the behavior of the normal decay length $\lambda_\text{max}$ 
(see Eq.~\eqref{eq:lambdamax} and Fig.~\ref{fig:6}), the interfacial
tension $\gamma$ levels off
for $L^*\gg2\pi$.}
\label{fig:8}
\end{figure}

The findings discussed so far lead to the picture of a surface layer of
thickness $\lambda_\text{max}$ in which a non-uniform surface charge
density, characterized by a lateral length scale $L$, can be sensed by the
electrolyte solution.
This thickness $\lambda_\text{max}$ is found to increase as function of $L$ as
long as $L^*= \kappa L \ll 2\pi$, whereas it is approximately constant for
$L^* \gg 2\pi$.
Therefore, one can expect that the interfacial tension $\gamma$ (see
Subsec.~\ref{sec:methods:interfacialtension}) exhibits the same trend.
This is indeed the case, as it is shown in Fig.~\ref{fig:8} for the
surface charges $\sigma^*=-3.3$ (red curve) and $-1.1$ (blue curve).
However, the interfacial tension $\gamma$ of a non-uniformly charged substrate
turns out to be limited to at most a few percent above the interfacial tension
$\gamma^{(1)}$ of a uniformly charged substrate with the same mean surface
charge density.


\section{\label{sec:conclusions}Summary, conclusions, and Outlook}

The present investigation is devoted to the structure formation in a dilute
electrolyte solution close to a non-uniformly charged planar substrate (see
Fig.~\ref{fig:1}).
In dilute electrolyte solutions the Debye screening length $\kappa^{-1}$ is
substantially larger than the size of the fluid molecules $R$ so that, in
principle, the spatial region of according thickness $\kappa^{-1}$
close to a charged substrate can be sensitive to the surface charge
distribution.
However, the lateral length scale $L$ of the charge distribution on the
substrate turns out to play a role, too.
In the present study periodic charge distributions with periodicity $L$ of
arbitrary magnitude are considered (see Fig.~\ref{fig:2}), and the
corresponding laterally non-uniform number density profiles of the fluid
particles are calculated via expansion about the profiles of a uniform
substrate with the same mean surface charge density (see Fig.~\ref{fig:3}).
It is found that the lateral structure is mainly determined by the
electrostatic potential, i.e., not by molecular-ranged forces like the
hard-core interaction, so that the laterally non-uniform contributions of the
number density profiles can be accurately approximated by a
Debye-H\"{u}ckel-like expression (see Eq.~\eqref{eq:dh3}), disregarding
hard-core contributions (see Fig.~\ref{fig:4}).
As a consequence, for normal distances not too close to the substrate, i.e.,
at $z$-coordinates with $\wt{\kappa}(z)/\kappa$ in Fig.~\ref{fig:5} close to
unity, the lateral contributions of the electrostatic potential, and hence of
the number densities, decay on the scale $\lambda_\text{max}$ given in
Eq.~\eqref{eq:lambdamax} (see Fig.~\ref{fig:6}).
For lateral length scales $L$ with $L^*=\kappa L \ll 2\pi$, the normal decay
length is varying with $L$ according to $\lambda_\text{max} \approx
L/(2\pi)$, whereas for $L^* \gg 2\pi$ it levels off at the value of the
Debye length, $\lambda_\text{max} \approx \kappa^{-1}$.
As shorter length scales $L^*\ll2\pi$ decay more rapidly than larger ones,
a washing out of fine details at increasing distance from the surface occurs
(see Fig.~\ref{fig:7}).
Ultimately only structures at length scales $L^*\gg2\pi$ contribute to the
lateral structure.
In terms of the interfacial tension of the non-uniformly charged substrate
an increase with $L$ is observed for $L^*\ll2\pi$ which saturates for
$L^*\gg2\pi$ (see Fig.~\ref{fig:8}).

Equation~\eqref{eq:psihat} in conjunction with Eq.~\eqref{eq:dh3} states,
that at distances $z\gg\lambda(q)$ (see Eq.~\eqref{eq:lambda}) details of
a surface charge distribution with wave number $q=|\vec{q}|$ become irrelevant
for the lateral structure of an adjacent electrolyte solution.
Hence, at larger distances from the substrate, only less fine details of a
surface charge distribution can be resolved.
Ultimately, at distances $z\gtrsim\kappa^{-1}$ details with wave numbers
$q=|\vec{q}|\gtrsim\kappa$, i.e.\ with lateral length scales $L\lesssim2\pi
\kappa^{-1}$, are washed out so that only surface structures with
lateral length scales $L\gtrsim2\pi\kappa^{-1}$ matter.
The strength of the influence of these large-scale structures decay
exponentially with a decay length given by the Debye screening length
$\kappa^{-1}$.
Therefore, when modeling electrolyte solutions with molecular length scale
$R$, one can safely ignore surface non-uniformities at length scales $L\lesssim
2 \pi R$, which, for molecular fluids, can be close to a nanometer.
Finally, the present study shows that macroscopic descriptions of electrolyte
solutions, i.e., on length scales larger than the Debye length $\kappa^{-1}$,
are carried out consistently by considering surface details on lateral
length scales larger than $2\pi\kappa^{-1}$ only.

Two main conclusions can be drawn from the present study:
(i) Microscopic hard-core interactions have negligible influence on the
lateral structure formation of electrolyte solutions close to non-uniformly
charged substrates.
(ii) Fine details of lateral non-uniformities have negligible influence
beyond a certain (short) distance from the substrate.
Accordingly, the approach of disregarding the size of molecules and
treating them as point particles (see many previous theoretical studies
concerning the interaction between non-uniformly charged colloidal particles)
can be justified or readily adjusted.
Generally, the present study shows that on macroscopic length scales only
macroscopically large features of the surface structure are visible.
This allows for local descriptions of fluids in terms of partial
differential equations, e.g., the Young-Laplace equation in hydrostatics.
The dominant correlation length of a fluid, which for a dilute electrolyte
solution of a non-critical solvent is the Debye length, separates length scales
into \emph{macroscopic} and \emph{microscopic} ones.
From a microscopic point of view, there is a smooth crossover of the fluid
structure from small to large length scales, whereas microsopic details
can be safely ignored from a macroscopic point of view.

Several directions of applications of the gained insight are conceivable:
The presented approach, i.e., to consider deviations from laterally
uniform reference density profiles and to ignore hard-core interactions, could
be exploited in various numerical analyses of fluid structures, including
computer simulations.
This way studies of large laterally non-uniform systems could become feasible.
Furthermore, given a certain length scale, the above insight is useful in
order to distinguish relevant from irrelevant surface details.
This is of importance not only for theoretical considerations or numerical
applications, but also for efficiently solving practical problems,
such as guiding flows in nanochannels, patterning surface structures of
catalytic reactors, or designing electrochemical devices.
Finally, a common understanding of the small effect microscopic features have
on macroscopic length scales (and vice versa) could be helpful for the
scientific discourse by avoiding confusion when comparing experimental or
theoretical results obtained within methods the spatial resolution of
which are associated with incompatible length scales.


\end{document}